\documentclass[usenatbib,fleqn,usedcolumn]{mnras}

\usepackage[british]{babel}
\usepackage{newtxtext}
\usepackage[slantedGreek]{newtxmath}

\let\la=\lesssim
\let\ga=\gtrsim

\usepackage[T1]{fontenc}
\usepackage{graphicx}
\usepackage{ae,aecompl}

\usepackage{amsmath}	

\usepackage{ctable}
\usepackage{url}
\usepackage{xspace}
\usepackage{pdflscape}

 \hypersetup{pdfauthor={C. C. Hayward, M. Sparre et al.},
               pdftitle={Submillimetre galaxies as constraints on feedback models},
               pdfkeywords={cosmology: theory -- methods: numerical -- galaxies: evolution -- galaxies: formation -- galaxies: star formation -- galaxies: starburst},
               bookmarksnumbered=true}

\newcommand{\cch}[1]{{#1}}

\newcommand\sref[1]{\hyperref[#1]{Section~\ref*{#1}}}
\newcommand\fref[1]{\hyperref[#1]{Fig.~\ref*{#1}}}
\newcommand\Eqref[1]{equation~(\hyperref[#1]{\ref*{#1}})}
\newcommand\tref[1]{\hyperref[#1]{Table~\ref*{#1}}}
\newcommand\msun{\text{M}_{\sun}\xspace}
\newcommand\msunperyr{\text{M}_{\sun}~\text{yr}^{-1}\xspace}
\newcommand{\mstar}{M_{\star}}
\newcommand{\mdust}{M_{\rm dust}}

\newcommand\rev[1]{#1}
\newcommand{\itng}{\textit{IllustrisTNG}\xspace}
\newcommand{\illustris}{\textit{Illustris}\xspace}

\title[SMGs as constraints on feedback models]{Submillimetre galaxies in cosmological hydrodynamical simulations -- an opportunity for constraining feedback models}

\author[Hayward, Sparre et al.]{
\parbox[t]{\textwidth}{
		Christopher C. Hayward$^{1}$\thanks{E-mail: \href{mailto:chayward@flatironinstitute.org}{chayward@flatironinstitute.org}},
		Martin Sparre$^{2,3,4}$, Scott C. Chapman$^5$, Lars Hernquist$^6$, Dylan Nelson$^7$, R\"udiger Pakmor$^{7}$, Annalisa Pillepich$^8$,
		Volker Springel$^7$, Paul Torrey$^{9}$, Mark Vogelsberger$^4$, and Rainer Weinberger$^6$
}
\vspace*{6pt} \\
$^1$Center for Computational Astrophysics, Flatiron Institute, 162 Fifth Avenue, New York, NY 10010, USA\\
$^2$Institut f\"ur Physik und Astronomie, Universit\"at Potsdam, Karl-Liebknecht-Str.\,24/25, 14476 Golm, Germany\\
$^3$Leibniz-Institut f\"ur Astrophysik Potsdam (AIP), An der Sternwarte 16, 14482 Potsdam, Germany\\
$^4$Department of Physics, Kavli Institute for Astrophysics and Space Research, Massachusetts Institute of Technology, \\
Cambridge, MA 02139, USA\\
$^5$Department of Physics and Atmospheric Science, Dalhousie University, 6310 Coburg Road, Halifax B3H 4R2, Canada\\
$^6$Harvard-Smithsonian Center for Astrophysics, 60 Garden Street, Cambridge, MA 02138, USA\\
$^7$Max-Planck-Institut f\"ur Astrophysik, Karl-Schwarzschild-Str. 1, D-85741 Garching, Germany\\
$^8$Max-Planck-Institut f\"ur Astronomie, K\"onigstuhl 17, D-69117, Heidelberg, Germany\\
$^9$Department of Astronomy, University of Florida, 211 Bryant Space Sciences Center, Gainesville, FL 32611, USA
}

\begin{document}


\pagerange{\pageref{firstpage}--\pageref{lastpage}} \pubyear{2020}
\maketitle

\label{firstpage}

\begin{abstract}
Submillimetre galaxies (SMGs) have long posed a challenge for theorists, and self-consistently reproducing the properties of the
SMG population in a large-volume cosmological hydrodynamical simulation has not yet been achieved.
\rev{We use a scaling relation derived from previous simulations plus radiative transfer
calculations} to predict the submm flux densities
of simulated SMGs drawn from cosmological simulations from the \illustris and \itng projects
\rev{based on the simulated galaxies' star formation rates (SFRs) and dust masses}
and compare the predicted number counts with observations.
We find that the predicted SMG number counts based on \itng are significantly less than observed
(more than 1 dex at $S_{850} \ga 4$ mJy). The simulation from the original \illustris project
yields more SMGs than \itng: the predicted counts are consistent with those observed at both $S_{850} \la 5$ mJy
and $S_{850} \ga 9$ mJy and only a factor of $\sim 2$ lower than observed at intermediate flux densities.
\rev{The redshift distribution of SMGs with $S_{850} > 3$ mJy in \itng is consistent with the observed distribution,
whereas the \illustris redshift distribution peaks at significantly lower redshift (1.5 vs. 2.8).}
We demonstrate that \itng hosts fewer SMGs
than \illustris because in the former, high-mass ($\mstar \sim 10^{11} \, \msun$) $z \sim 2-3$ galaxies have lower dust masses and
SFRs than in \illustris owing to differences in the sub-grid models for stellar or/and active galactic
nucleus (AGN) feedback between the two simulations (we unfortunately cannot isolate the specific cause(s) post hoc).
Our results demonstrate that because our method enables predicting SMG number counts in post-processing with a negligible computational expense,
SMGs can provide useful constraints for tuning sub-grid models in future large-volume cosmological simulations.
\end{abstract}
\begin{keywords}
cosmology: theory -- methods: numerical -- galaxies: evolution -- galaxies: formation -- galaxies: star formation -- galaxies: starburst.
\end{keywords}

\section{Introduction} \label{sec:introduction}

Submillimetre galaxies (SMGs) are a class of rapidly star-forming (SFR $\sim 10^2-10^3 \, \msunperyr$), highly dust-obscured galaxies
that were discovered in the late 1990s (\citealt{Smail:1997,Barger:1998,Hughes:1998,Eales:1999}; see \citealt{Casey2014} for a recent review). Since that time, they have been
the focus of much observational work for multiple reasons. For example, they have inferred SFRs that are unprecedented in the local Universe
(a `typical' SMG has an SFR a factor of a few greater than Arp 220, and some have SFRs over an order of magnitude greater).
Their SFR surface densities are sufficiently high that they approach the theoretical maximum SFR surface density at which
momentum injection from stellar feedback can unbind the remaining gas and thus halt star formation \citep{Scoville:2003,Murray:2005,Thompson:2005}.
They are estimated to contribute
as much as half of the SFR density of the Universe at $z \sim 2-3$ \citep{Chapman:2005,Dunlop2017,Michalowski2017,Smith2017}.
Their number densities ($N \sim 10^{-5} - 10^{-6}$ cMpc$^{-3}$ at $z \sim 2-3$) and stellar (median $\mstar \sim 10^{11} \, \msun$; e.g.
\citealt{daCunha2015,Ma2015spt,Miettinen2017sed}) and halo ($M_{\mathrm{h}} \sim 10^{13} \, \msun$; \citealt{Blain:2004,Hickox:2012,Wilkinson2017})
masses suggest that they may be the progenitors of compact quiescent galaxies \citep{Toft2014}.
SMGs at $z \ga 3$ serve as beacons of the highest-mass dark matter halos at a given epoch \citep{Miller2015,Miller2020}, potentially enabling
tests of $\Lambdaup$CDM \citep{Marrone2018}.
Most importantly for the present work, reproducing the observed SMG number counts and redshift distribution has posed a challenge for theorists since
their discovery \citep{Baugh:2005,Primack:2012}, and their physical nature is still vigorously debated.

Modelling the SMG population self-consistently is extremely challenging. Because of their rarity, large volumes must be simulated. However,
one must also resolve individual galaxies sufficiently well to capture physical processes that may be responsible for their high SFRs, including
the classical explanation, merger-induced starbursts \citep[e.g.][]{Engel:2010,Chakrabarti:2008SMG,Narayanan:2010smg,H11}, and disc instabilities
\citep{Lacey2015}. Moreover, to accurately model SMGs, one must predict submm fluxes of simulated galaxies because there is not a one-to-one relation between
submm flux density and SFR, mainly owing to the dependence of the dust temperature distribution and thus submm flux density on the dust mass in
addition to the SFR \citep{H11,
Safarzadeh2016ir_seds}. For this purpose, it is necessary to perform dust radiative transfer, the results of which depend on the distribution of dust and sources of
emission (stars and active galactic nuclei, AGN) on sub-galactic scales.

For the above reasons, hybrid approaches -- rather than large-volume cosmological simulations with full radiative transfer -- have typically been
employed to model the SMG population.
\rev{The earliest theoretical studies about SMGs employed semi-analytic models, in which submm fluxes were predicted via simplified radiative transfer
calculations assuming an idealised axisymmetric
geometry \citep[e.g.][]{Granato:2000,Baugh:2005} or using an analytic model for dust attenuation and empirical SED templates \citep[e.g.][]{Somerville:2012}.}
\citet{Granato:2000} compared the SMG number counts and redshift distribution predicted by the {\sc galform} \citep{Cole:2000} SAM
with that observed. They found that the model under-predicted the observed counts by a factor of $\sim 20-30$.
Other SAMs \citep{Fontanot:2007,Somerville:2012} also significantly under-predicted the bright SMG number counts.
\citet{Baugh:2005} explored various modifications to the {\sc galform} model, searching for a means to match the SMG number counts and redshift distribution
while not violating other observational constraints, such as the $z = 0$ stellar mass function. They found that by assuming a flat stellar initial mass function
(IMF) in starbursts (specifically, $dn/d\log m = $ constant for the mass range $0.15 < m/\msun < 125$),
in addition to other subdominant modifications to the model, they were able to reproduce the SMG number counts and redshift distribution.
In the \citet{Baugh:2005} model, SMGs are predominantly minor merger-driven starbursts \citep{Gonzalez:2011}.
However, due to the many free parameters employed in SAMs, it is possible to find multiple qualitatively distinct solutions that satisfy observational constraints
equally well \citep[e.g.][]{Lu:2011a,Henriques2013}, so it is unclear that a top-heavy IMF is necessary, rather than simply sufficient given the other components
of the model, to reproduce the SMG population. Indeed,
in the current version of the {\sc galform} model \citep{Lacey2015}, the IMF assumed for starbursts, $dn/d \log m \propto m^{-1}$, is still top-heavy compared
to the \citet{Kroupa:2001} IMF ($dn/d\log m \propto m^{-1.3}$ for $m > 0.5 ~\msun$) but considerably steeper
than that assumed by \citet{Baugh:2005}, and SMGs are predominantly starbursts driven by disc instabilities rather than mergers \citep{Cowley2015a,Lacey2015}. Moreover,
\citet{Safarzadeh2017} demonstrated that the SMG number counts and redshift distribution predicted based on the \citet{Lu:2014cy} SAM, which uses a
standard \citet{Chabrier:2003} IMF, agree reasonably well with observations.

\rev{Some studies have employed approaches based on
performing post-processing dust radiative transfer on idealised (non-cosmological) galaxy merger simulations.} \citet[][see also \citealt{Chakrabarti:2008SMG}; \citealt{H11,H12}]{Narayanan:2010smg}
demonstrated that merger-induced starbursts could produce submm flux densities as high as observed. \citet{HN13} combined the results of isolated disc and galaxy merger
simulations with a semi-empirical model for the redshift evolution of galaxy mass functions, merger rates, gas fractions and other properties to predict the SMG
number counts and redshift distribution, finding good agreement with observations despite employing a standard \citet{Kroupa:2001} IMF.
In their model, SMGs are a mix of isolated `quiescently' star-forming disc galaxies, pre-coalescence mergers blended into a single submm source (see below), and
merger coalescence-induced starbursts.
Although the hybrid approach employed by \citet{HN13} satisfies various observational constraints, such as the $z = 0$ stellar mass function, by construction,
one significant downside is that it effectively sidesteps some important issues, such as self-consistently predicting the correct $z \sim 2-3$ stellar mass function, and is far from `ab initio'.

\citet{Narayanan2015} performed radiative transfer in post-processing on a zoom simulation of a halo with a $z = 2$ halo mass of $3 \times 10^{13} \, \msun$.
The central galaxy reached a peak submm flux density of $S_{850} \sim 25$ mJy, comparable to the flux densities of the brightest SMGs known \citep{Hayward:2013limits}.
\citet{Narayanan2015} argued that major mergers did not play an important role in driving the galaxy's high submm flux density. They posited that SMGs are typically massive
galaxies undergoing minor mergers and that stellar feedback -- which results in bursts of star formation, subsequent outflows, and re-accretion in a
`galactic fountain' \citep{Muratov2015,Sparre2017,HH17} -- `saves' gas accreted by the halo for consumption at $z \sim 2-3$, thus resulting in an SMG
phase consistent with the peak of the observed SMG redshift distribution.
However, as with idealised, non-cosmological simulations, zoom simulations suffer from a lack of statistics; thus, \citet{Narayanan2015} did not present predictions for the
SMG number counts and redshift distribution, and it may not be possible to infer the nature of a `typical' SMG from the evolution of a single halo.

Despite the aforementioned difficulties in using large-volume cosmological hydrodynamical simulations to model the SMG population, there have been a few
such works published to date.
\citet{Dave:2010} performed a hydrodynamical simulation with a volume of $(96 h^{-1} \, \mathrm{Mpc})^3$. Given an
estimate for the number density of $z \sim 2$ SMGs ($1.5 \times 10^{-5} \, \mathrm{cMpc}^{-3}$),
they computed the expected number of SMGs that the simulation should contain, 41, and analysed the 41
most rapidly star-forming galaxies in the $z = 2$ snapshot (effectively assuming a monotonically increasing relationship between submm flux
density and SFR). In the \citet{Dave:2010} model, SMGs correspond to the high-mass $(\mstar \sim 10^{11-11.7} \, \msun)$
end of the star-forming galaxy `main sequence', and their star formation is typically not driven by major mergers.
The properties of the simulated galaxies agreed well with some of the observed properties of SMGs, but the SFRs were lower than
those inferred for real SMGs by a factor of $\sim 3$. \citet{Dave:2010} noted that one possible explanation for this discrepancy is that the observationally
inferred SFRs, which rely on the assumption of a standard Milky Way-type IMF, would overestimate the true SFR of SMGs if the IMF in SMGs were top-heavy,
as argued in some of the works mentioned above. However, we note that at $z = 2$, at fixed stellar mass, \emph{all} galaxies in the \citet{Dave:2010}
simulation, not only the SMG analogues, have SFRs lower than observed (i.e.~the normalisation of the $z \sim 2$
SFR--stellar mass relation in the simulation is systematically
below that observed). Consequently, reconciling the SFRs of the observed and simulated galaxies would require a top-heavy IMF in lower-mass,
more-typical galaxies in addition to SMGs.

\citet*{Shimizu:2012} performed a cosmological hydrodynamical simulation with a comoving volume of $(100 h^{-1} \, \mathrm{Mpc})^3$.
They employed a simple model to predict the submm flux densities of their simulated galaxies: assuming a spherically symmetric dust geometry, they
computed the total luminosity absorbed by dust. Then, assuming dust of a single temperature in thermal equilibrium and neglecting dust
self-absorption, they computed the submm flux densities of their simulated galaxies. Their predicted counts agree well with those observed
for $S_{850} \ga 1$ mJy, but they under-predict the counts of fainter sources. However, this success should be taken with a grain of salt,
as their simplified model for computing the submm flux density may over-predict the submm flux density: applying the same model
(i.e.~using the $S_{850}(\mathrm{SFR}, \mdust)$ scaling for an optically thin single-temperature modified blackbody) to hydrodynamical
simulations of isolated disc galaxies and galaxy mergers on which they performed radiative transfer, \citet{H11} found that
the simplified model over-predicts the flux densities (relative to the results of the radiative transfer calculations) by $\sim 0.3-0.5$ dex.
Because of the steepness of the bright end of the SMG number counts, reducing the predicted fluxes of their model SMGs by this
seemingly modest factor would cause the number counts to be significantly under-predicted.

\citet{McAlpine2019} compared the properties of `submm-bright' galaxies (which they define by $S_{850} \ge 1$ mJy) in the {\sc eagle}
simulation \citep{Schaye2015} with those of observed SMGs. The submm fluxes were computed by performing dust radiative transfer
directly on the galaxies from the simulation. Notably, the redshift distribution of sources with $S_{850} \ge 1$ mJy agrees
reasonably well with that observed, \rev{but the predicted number counts of sources with $S_{850} \ge 3$ mJy are more
than two orders of magnitude less than those observed (fig. 9 of \citealt{Cowley2019}; see also \citealt{Wang2019})}.

\rev{\citet{Lovell2020} performed dust radiative transfer on galaxies from {\sc simba} \citep{Dave2019}. Their predicted counts are
closer to those observed than are those for {\sc eagle} or, as we shall see below, \itng, but they are still a factor of a few to $\sim 10$
less than observed, and the SMG redshift distribution is skewed toward somewhat higher redshift than observed.}

\rev{The main advantage of our work \rev{(and that of \citealt{McAlpine2019} and \citealt{Lovell2020})}
relative to the cosmological-simulation-based SMG studies of \citet{Dave:2010} and \citet{Shimizu:2012} is
how we define SMGs.}
For this purpose, it is necessary to compute the submm flux density directly because both the SFR and dust mass
affect the FIR/submm SED \citep{H11,Lanz2014,Safarzadeh2016ir_seds}.
For this reason, a galaxy with a high SFR but low dust mass may actually have a lower submm flux density than
one with a lower SFR but high dust mass, owing to the former galaxy having a `hotter' SED. Consequently, the monotonic mapping between submm
flux density and SFR assumed in \citet{Dave:2010} does not hold in detail \citep{H11,HN13}.
\rev{Since the mass and spatial resolution state-of-the-art large-volume cosmological hydrodynamical simulations are possibly} still insufficient to perform
dust radiative transfer directly on the simulated galaxies (i.e.\,without resampling in both space and time; \citealt{Trayford2017}), \rev{and
to avoid the significant computational expense associated with doing so,
we instead} assign submm flux densities to our simulated galaxies using a fitting function that gives the submm flux density as a function of SFR and dust mass.
The fitting function was derived from the results of performing dust radiative transfer on hydrodynamical simulations of isolated disc and merging galaxies
\citep{H11,HN13}; \cch{see \sref{sec:methods} for details}. As discussed above, this approach is superior to that of \citet{Shimizu:2012} because the single-temperature, optically thin modified blackbody scaling
assumed in that work tends to over-predict the submm flux densities of the simulated galaxies by a factor of $\sim 0.3-0.5$ dex \citep{H11}, implying that the
SMG number counts presented in \citet{Shimizu:2012} are significantly higher than they would be had a more accurate method for computing submm flux
densities been used.

One issue that has complicated previous comparisons between theoretical models and observations is blending of multiple SMGs into
a single submm source. Models predict that blending of both early-stage mergers \citep{H11,H12,HN13} and physically unassociated galaxies
\citep{HB13,MunozArancibia2015,Cowley2015a} should result in a significant fraction of observed single-dish submm sources
breaking up into multiple sources when observed at higher resolution with e.g. ALMA, as has been observed \citep[e.g.][]{Karim2013,
Hodge2013,Simpson2015,H18}. Blending causes the number counts derived from single-dish observations (e.g. those based on SCUBA-2 maps) to exceed the true
SMG number counts, as demonstrated explicitly by the comparison of LABOCA and ALMA counts shown in \citet{Karim2013},
although the magnitude of this effect is still debated. For simplicity,
we do not treat blending of submm sources in this work, so we will compare the simulated SMGs only with ALMA observations of spatially resolved SMGs.

\cch{In this work, we show that the SMG number counts predicted from \illustris using our simple method are in reasonable agreement with observations,
whereas those predicted from \itng are significantly less than observed. However, it has been shown in numerous previous works that \itng better
reproduces various aspects of the general galaxy population than \illustris, and thus the aforementioned result should not be interpreted as an
indication that \illustris is more correct than \itng. Instead, our results indicate a need for further revision of the sub-grid models and 
demonstrate the utility of SMG number counts for constraining said
models for future large-volume cosmological simulations, especially given that our method for predicting the SMG counts incurs negligible computational expense.}

The remainder of this work is organised as follows: in \sref{sec:methods}, we summarise the \illustris and \itng simulations and the method for
generating mock SMG catalogues from these simulations. \sref{sec:results} compares the predicted number counts
with observations. In \sref{sec:discussion}, we outline how our method for
predicting SMG number counts from large-volume cosmological simulations can be used to constrain sub-grid feedback models,
\rev{discuss how the quenched fractions of massive $z \sim 2-3$ galaxies in the simulations compare with observational constraints,}
and detail some limitations and future work. In \sref{sec:conclusions}, we conclude.

\section{Methods} \label{sec:methods}

\subsection{The \illustris and \itng simulations} \label{sec:sims}

Here, we summarise the details of the large-volume cosmological simulations from the \illustris and \itng projects used in this work.
Both simulations were run using {\sc arepo} \citep{Springel:2010arepo}, an unstructured moving-mesh hydrodynamics code.
The details of the physical model employed in \illustris are presented in \citet[][see also \citealt{Vogelsberger2013} and \citealt{Torrey2014}]{Vogelsberger2014illustris},
so we will only briefly list the key aspects here. The models for star formation and stellar feedback are based on \citet{Springel:2003}.
Star formation is implemented by stochastically spawning stellar particles at a rate set by a volume density-dependent Kennicutt-Schmidt law \citep{Kennicutt:1998,Schmidt:1959};
a density threshold of 0.13 cm$^{-3}$ is employed.
An effective equation of state is used to approximate how SNe heat the ISM, and stellar feedback-driven winds are implemented by stochastically kicking
and temporarily hydrodynamically decoupling gas cells in a manner intended to capture bipolar winds \citep{Springel:2003}. Importantly for the present work, stellar evolution, chemical enrichment, and gas return
are tracked. BH accretion is modelled as Eddington-limited modified Bondi-Hoyle accretion with a boost factor to account for the gas density near the BH being underestimated
due to resolution, and a two-stage model for AGN feedback, including both quasar and radio modes, is employed \citep{Springel:2005feedback,Sijacki:2007}.
\rev{The sub-grid models were tuned to reproduce the SFR density as a function of redshift and $z = 0$ SFR stellar mass--halo mass and black hole mass--stellar mass relations.}
Further details about and first results from the \illustris simulations are presented in \citet{Vogelsberger2014nature,Vogelsberger2014illustris,Genel2014illustris,Sijacki2015illustris}.
In this work, we use the \textit{Illustris-1} simulation, which has a volume of (106.5 Mpc)$^3$, $2 \times 1820^3$ resolution elements, a dark matter (baryonic) mass resolution of $6.26 ~(1.26) \times 10^6 \, \msun$, and
a $z = 0$ gravitational softening of 710 pc.

The \itng simulation that we employ, {\sc tng100}, has the same initial conditions \cch{(modulo minor differences
in the cosmology)} and resolution as \textit{Illustris-1}, but improvements to the numerical method and an updated physical model intended to alleviate
many of the discrepancies between the \illustris results and observations are employed.
\cch{We note that the aforementioned discrepancies refer to the overall galaxy population, such as the $z = 0$ galaxy colour bimodality; the properties of SMGs,
which represent an `extreme' sub-population of galaxies, were not considered.}
The updated numerical method and physical model are described in \citet{Pillepich2018itng} and \citet{Weinberger2017}, and further details
about the simulations and first results are presented in various works \citep[e.g.][]{Springel2018itng,Pillepich2018itngresults,Nelson2018itng,Naiman2018itng}.
\rev{The \itng model was tuned to reproduce the SFR density as a function of redshift and $z = 0$ galaxy stellar mass and stellar mass--halo mass relations, and
the BH mass--stellar mass relation, halo gas fractions and stellar half-mass radii of galaxies were also considered; see section 3.2 of \citet{Pillepich2018itng}.}
We refer the interested reader to those works for full details, and we will only highlight the modifications most relevant to the present work here. Regarding the AGN feedback model,
the low-accretion state mode was changed to a kinetic model \citep{Weinberger2017}. When computing the Bondi-Hoyle accretion rate, a boost factor is no longer employed; to prevent
too-slow early BH growth, a larger seed mass ($8 \times 10^5 h^{-1} \, \msun$, rather than the value of $10^{-5} h^{-1} \, \msun$ used in \illustris) is assumed. A few aspects of the model for
stellar feedback-driven winds were also changed \citep{Pillepich2018itng}. First, the wind calls are kicked isotropically rather than in a bipolar manner. A redshift dependence and floor for the wind velocity
are also employed. Also, unlike in \illustris, a fraction of the wind energy is deposited as thermal (rather than kinetic) energy \citep{Marinacci2014}, and a metallicity dependence of
the mass loading factor, such that winds are less effective at higher metallicity, is assumed.

\subsection{Generating simulated SMG catalogues}

As discussed in \sref{sec:introduction}, to identify SMGs in the simulation, it is ideal to compute individual simulated galaxies' submm flux densities via dust radiative transfer.
The mass and spatial resolution of \illustris and \itng are insufficient to resolve the sub-kpc structure of the ISM, and owing to both the sub-grid equation of state treatment of stellar feedback
and resolution, such simulations tend to feature overly puffy discs and thus underestimate the attenuation \citep{Trayford2017}. Consequently, we do not perform radiative transfer directly.
Instead, we use the following relation based on the results of performing radiative transfer on higher-resolution hydrodynamical simulations of idealised isolated discs and mergers \citep{H11,HN13}:
\begin{equation}\label{eq:s850}
\frac{S_{850}}{\left[\text{mJy}\right]} = 0.81 \left( \frac{\text{SFR}}{100 \, \left[\text{M}_\odot \,\text{yr}^{-1}\right]} \right)^{0.43} \left( \frac{\mdust}{10^8 \, \left[\text{M}_\odot\right]} \right)^{0.54},
\end{equation}
where \cch{SFR is the `instantaneous' SFR associated with the star-forming gas cells (for consistency with \citealt{H11})} and $\mdust$ is the dust mass.
When applying the relation, following \citet{HB13}, we incorporate \cch{a scatter of 0.13 dex
(independent of SFR and dust mass)}
assuming a Gaussian distribution \citep{H11}.
The submm flux densities predicted using this relation also agree well with the results of performing dust radiative transfer directly on cosmological zoom simulations
\citep{Liang2018,Cochrane2019} and a semi-analytic model (A.\,Benson, private communication).
Moreover, by combining the above with the observed SFR-$\mstar$ and $\mdust(\mstar,z)$ relations, one can derive an $S_{850}-\mstar$ relation
\citep{HN13,Hayward:2012thesis} that agrees reasonably well with that observed for SMGs \citep{Davies:2013}.
\cch{Although given sufficient resolution and unlimited computing time, it would be preferred to perform radiative transfer directly, our approach has the advantage of trivial
computational expense, and it may actually be more accurate than direct radiative transfer given the resolution of state-of-the-art large-volume simulations such as \illustris and \itng.}

In the above, we approximate the total SFR and ISM metal mass of a simulated galaxy by summing over the values for all cells within 25 kpc of the subhalo centre, excluding those that are
contained in subhaloes of the halo under consideration (i.e.\,those in satellite galaxies; recall that we will compare with observations of resolved SMGs, in which satellite galaxies should typically not be
blended with central galaxies). We have confirmed that the results are insensitive to the radial cut employed as long as it is $\ga 10$ kpc.
Since dust is likely destroyed in hot halo gas \citep[e.g.][]{McKinnon2016,McKinnon2017,Popping2017}, when computing the ISM metal mass, we only consider `ISM gas' defined
using the temperature-density cut from \citet{Torrey:2012disks}:
\begin{equation}
\log_{10} \left(\frac{T}{[K]}\right) < 6 + 0.25 \log_{10} \left( \frac{\rho}{10^{10} \, \left[\msun \, h^2 \, \mathrm{kpc}^{-3}\right]} \right).
\end{equation}
Given this value, we then compute the dust mass using a dust-to-metal ratio of 0.4 \citep{Dwek:1998,James:2002}.
The results are not strongly dependent on the exact dust-to-metal ratio used because of the sub-linear scaling between submm flux density and dust mass \cch{(e.g. using a value of 0.5 would boost
the submm flux density by $\sim 13$ per cent).}

\subsection{Number counts}

The process described in the previous subsection yields a submm (850-\micron) flux density, $S_{850}$, for each subhalo in the simulation (in addition to other important properties of each subhalo, including the SFR; stellar, gas, metal, and dust
masses; and redshift). Given the SMG catalogues derived from each time snapshot of the simulations, we then compute the number counts and redshift distribution.
To determine the cumulative number counts of galaxies, $dN(>S_{850})/d \Omega$, we first calculate the number of sources with a given submm flux per unit comoving volume for each snapshot. Next,
we multiply the number density at a given redshift by the redshift-dependent comoving volume element and integrate over redshift \citep[see equation 3 in][]{HN13}.

\section{Results} \label{sec:results}

\begin{figure}
\centering
\includegraphics[width = 0.48 \textwidth]{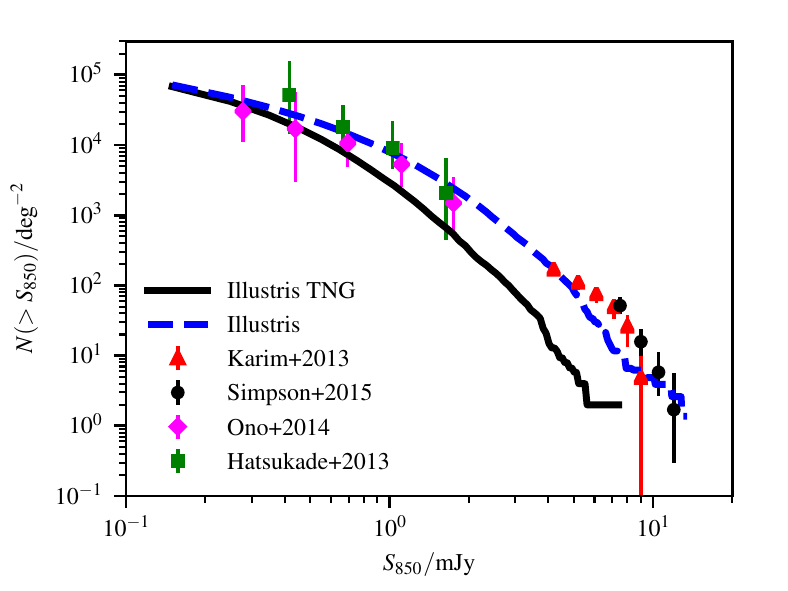}
\caption{Cumulative 850-\micron ~number counts of SMGs predicted using \itng (\emph{solid black line}) and \illustris (\emph{dashed blue line}).
\rev{Because we predict the counts of individual SMGs rather than submm source counts for some fixed resolution (i.e.\,we do not treat blending),} we compare only with number counts derived from ALMA observations:
\citet[][\emph{red triangles}]{Karim2013}, \citet[][\emph{black circles}]{Simpson2015},
\citet[][\emph{green squares}]{Hatsukade2013}, and \citet[][\emph{magenta diamonds}]{Ono2014}.
The \itng counts are significantly lower than those observed: for $S_{850} \ga 4$ mJy, the counts are under-predicted by 1 dex or more.
The \illustris-based counts are in excellent agreement with the observed counts at $S_{850} \la 5$ mJy and $S_{850} \ga 9$ mJy, and they are only a factor of $\sim 2$ lower at intermediate flux densities.}
\label{fig:num_cts}
\end{figure}

\fref{fig:num_cts} compares the 850-\micron ~number counts predicted using \itng and \illustris with those observed. As explained in \sref{sec:introduction}, we do not treat the effects of blending due to the
coarse angular resolution of single-dish submm telescopes, so we only compare with counts based on interferometric observations with ALMA since such observations have sufficient angular resolution
to resolve multi-component sources into individual SMGs \citep[e.g.][]{Karim2013}. The \itng counts are significantly lower than observed: at $S_{850} \ga 4$ mJy, the observed counts are an order of magnitude
or more higher.
There are no SMGs in \itng with $S_{850} \ga 8$ mJy. The counts predicted using \illustris are in better agreement with those observed: they are consistent at $S_{850} \la 5$ mJy and $S_{850} \ga 9$ mJy,
and they are only a factor of $\sim 2$ lower than observed at intermediate flux densities.

\begin{figure}
\centering
\includegraphics[width = 0.48 \textwidth]{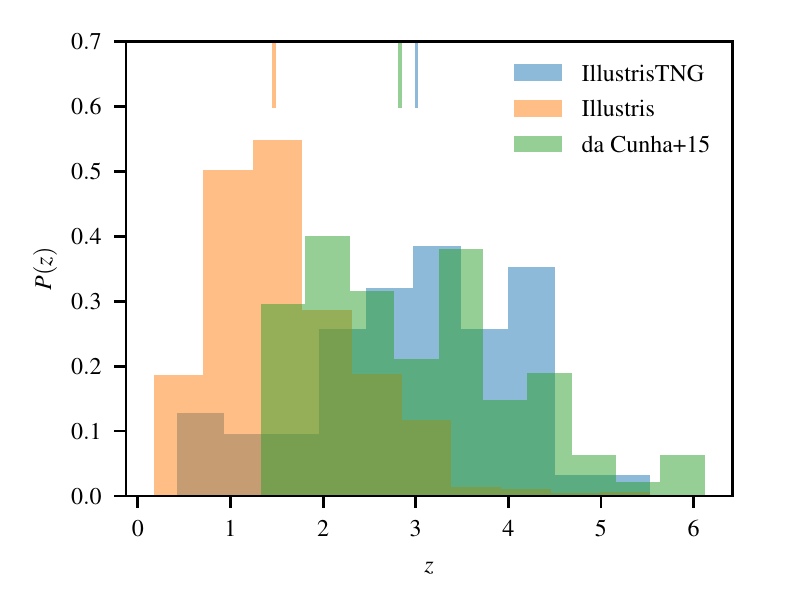}
\caption{\rev{Normalised redshift distributions for the simulated SMGs with $S_{850} \ge 3$ mJy and the observed SMGs from \citet{daCunha2015}. The median values
are shown by the vertical bars at the top of the plot. The \itng redshift distribution is similar to that of the observed sample,
whereas the distribution for \illustris peaks at significantly lower redshift than observed
($z \sim 1.5$ for \illustris vs. $z \sim 2.8$ for the \citealt{daCunha2015} sample).
}}
\label{fig:z_dist}
\end{figure}

\rev{\fref{fig:z_dist} compares the redshift distributions of SMGs with $S_{850} > 3$ mJy in the two simulations with the observed distribution for the ALESS sample \citep{daCunha2015}.
In contrast with the number counts, for the redshift distribution, \itng is in much better agreement with observations. The \itng and observed redshift distributions are broadly consistent,
with similar median redshifts (3.0 vs. 2.8). The \illustris redshift distribution peaks at a much lower redshift, 1.5, and differs dramatically from that observed. It is interesting that \itng
matches the observed redshift distribution significantly better than \illustris, but given that \itng underpredicts the number counts by more than an order of magnitude, it is unclear
whether this agreement is meaningful or purely coincidental.}

\begin{figure}
\centering
\includegraphics[width = 0.48 \textwidth]{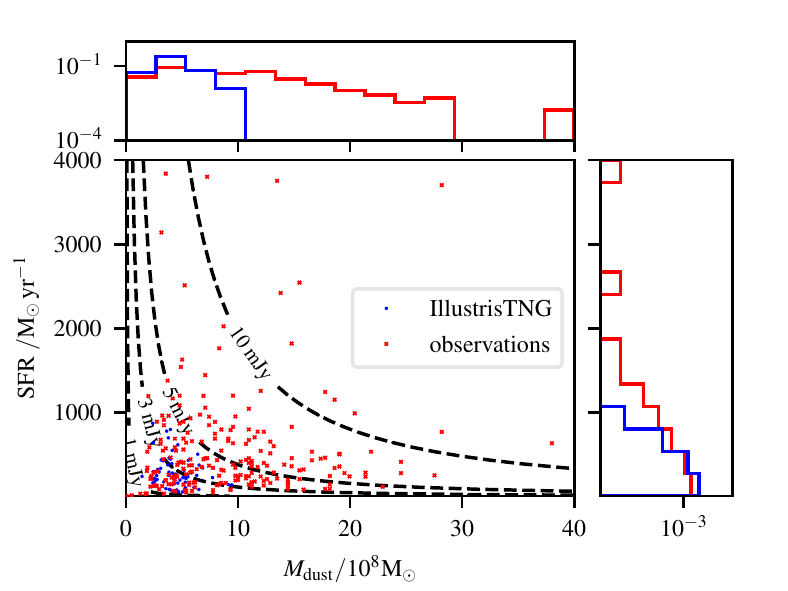}
\includegraphics[width = 0.48 \textwidth]{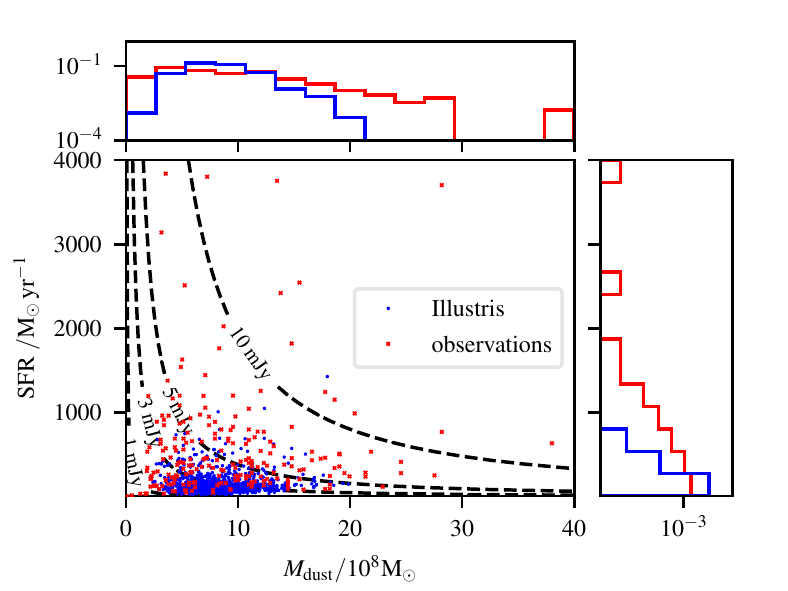}
\caption{Distribution of simulated SMGs with $S_{850} > 3$ mJy in the SFR--$\mdust$ plane (\emph{blue circles}) for \itng (\emph{top panel}) and \illustris (\emph{bottom panel}). The \emph{black dashed lines}
are contours of constant submm flux density ($S_{850} = $ 1, 3, 5, and 10 mJy) calculated using Eq.\,(\ref{eq:s850}) assuming $z=2$. Observed SMGs from \citet[][]{daCunha2015}
and \citet[][]{Miettinen2017sed}
are shown as \emph{red x's} for comparison. \rev{The probability density functions for the simulated and observed SMGs are shown in the marginal plots.} 
The simulated galaxies from \illustris span much of the SFR--$\mdust$ region occupied by the observed SMGs,
whereas both the $\mdust$ and, to a lesser extent, SFR values of the \itng galaxies are significantly lower than those of the observed bright SMGs.
}
\label{sfr_mdust}
\end{figure}

To understand the underlying reason(s) for the discrepancies between the counts predicted using \itng and \illustris, in \fref{sfr_mdust}, we show the distribution of the mock SMGs in the
SFR-$\mdust$ plane (blue circles) for \itng (top panel) and \illustris (bottom panel). Only simulated galaxies with $S_{850} > 3$ mJy are plotted.
The dashed lines show contours for various $S_{850}$ values assuming $z=2$; recall that in our model, the simulated
galaxies' SFRs, dust masses, and redshifts fully determine their submm flux densities (see Eq.\,\ref{eq:s850}).\footnote{\rev{Because we include scatter when computing
$S_{850}$ using Eq.\,(\ref{eq:s850}), many of the simulated SMGs with assigned $S_{850} > 3$ mJy lie on the lower-flux side of the $S_{850} = 3$ mJy contour.}}
The other points represent observed SMGs from \citet{daCunha2015}
and \citet{Miettinen2017sed}.\footnote{One SMG from \citet{Miettinen2017sed} with SFR $= 6501 \, \msunperyr$ and $\mdust = 1.7 \times 10^8 \, \msun$ has been
omitted to avoid excess white space.}
The SFR and $\mdust$ values for the observed SMGs were obtained by fitting their UV--mm spectral energy distributions (SEDs) with the {\sc magphys}
SED modelling code, which has been demonstrated to be able to accurately recover these and other galaxy properties by applying the code to simulated galaxy SEDs, for which the
`ground truth' is known \citep{Michalowski2012,HS15,SH15,SH18}.
Because we plot the simulated SMGs taken from all redshift snapshots rather than the results for a lightcone
covering an area equal to that probed by the observations, the densities of the simulated and observed points in a given region should not be directly compared, but the plots are
useful for determining whether the simulations and observations span the same range of SFR and $\mdust$ values. Moreover, because \illustris and \itng have identical volumes
(and initial conditions \cch{modulo minor differences in the cosmology}), directly comparing the density of points for the simulations in the two panels is meaningful.

The most significant difference between \itng and \illustris is that simulated galaxies in the former have significantly
lower dust masses, with no \itng galaxies having $\mdust \ga 9 \times 10^8 \, \msun$. In contrast, \illustris contains many simulated galaxies with dust masses greater this value.
\cch{For reference, the observed SMGs in the ALESS sample (for which the number counts
from \citealt{Karim2013} are plotted as red dots in \fref{fig:num_cts}) have a median dust mass
of $5.6 \times 10^8 \msun$, and the 16th-84th-percentile range of the likelihood distribution is
$(0.22-11) \times 10^9 \msun$ \citep{daCunha2015}.}
Some galaxies in \illustris have dust masses that are a factor of two higher than that of the most dust-rich galaxy in \itng, \cch{and as shown below, $z \approx 2$ galaxies with
$\mstar \approx 10^{11} \msun$ (i.e. `typical SMGs' in the simulations)
in \itng have dust masses that are typically a factor of 3 lower than those of their \illustris counterparts.}
Since in Eq. \ref{eq:s850}, $S_{850} \propto \mdust^{0.54}$, a factor of \cch{three} higher dust mass corresponds to a \cch{$\sim 80$} per cent higher submm flux density.
Given the steepness of the number counts, the relatively modest boost in flux density due to the higher dust masses of the \illustris galaxies
makes the \illustris counts agree much better with observations (\fref{fig:num_cts}).

The \itng galaxies also exhibit lower SFRs, although the difference in the SFR distributions of \illustris and \itng is less significant than for $\mdust$.
In \itng, 16 $z = 2$ galaxies have SFR $>500 \, \msunperyr$, and the maximum SFR reached is $896 \, \msunperyr$. \illustris has 29 $z = 2$ galaxies with SFR $>500 \, \msunperyr$,
and the most rapidly star-forming galaxy has SFR $= 1429 \, \msunperyr$.

\begin{figure}
\centering
\includegraphics[width = 0.48 \textwidth]{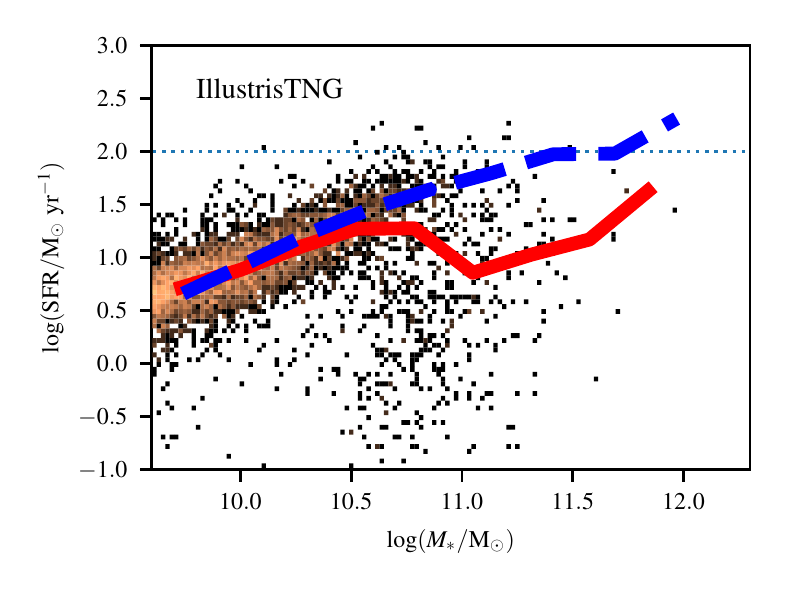}
\includegraphics[width = 0.48 \textwidth]{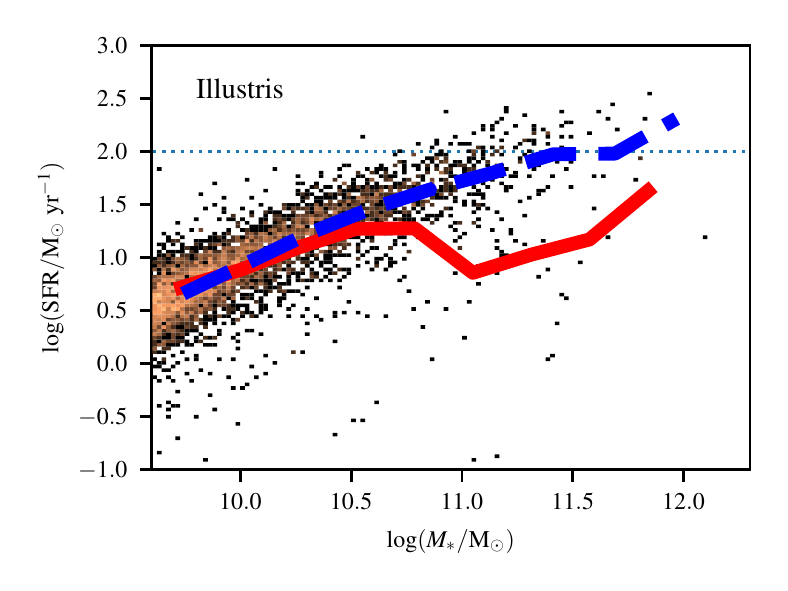}
\caption{SFR vs. stellar mass for \itng (\emph{top panel}) and \illustris (\emph{bottom}) at $z = 2$. All simulated galaxies with $\mstar \ga 10^{9.5} \, \msun$
are plotted (\rev{no SFR cut has been applied}). \rev{The red solid (blue dashed) lines show the running median SFR for \itng (\illustris).}
The horizontal lines are intended only to guide the eye.
The SFR--$\mstar$ relations of \itng and \illustris are similar for $\mstar
\la 10^{10.5} \, \msun$,
but at higher masses, the \itng galaxies have lower SFRs and a greater quenched fraction.}
\label{fig:sfr-mstar}
\end{figure}

In \fref{sfr_mdust}, we only considered simulated galaxies with $S_{850} > 3$ mJy.
We will now compare some key properties of all massive
galaxies in \illustris and \itng, which will further highlight the reasons for the differences between the \illustris and \itng SMG number counts.
\cch{We focus on $z \sim 2$ for this analysis because the SMG redshift distribution peaks around that redshift \citep[e.g.][]{Chapman:2005}.}
\fref{fig:sfr-mstar} shows the distribution of all \itng (top panel) and \illustris (bottom) galaxies in the $z = 2$ snapshot with $\mstar \ga 10^{9.5} \, \msun$
in the SFR--$\mstar$ plane. At $\mstar \la 10^{10.5} \, \msun$, both simulations feature a tight `star formation main sequence'
(SFMS; \citealt{Noeske:2007a,Daddi:2007}; \cch{see \citealt{Sparre2015} and \citealt{Donnari2019}
for detailed analyses of the SFR--$\mstar$ relations in \illustris and \itng, respectively}), and the typical SFR values at fixed stellar masses are similar.
At higher masses, many of the simulated galaxies are `quenched' (i.e.~lie significantly below the main sequence),
with a greater fraction of quenched galaxies in \itng than in \illustris;
\rev{how the quenched fractions in the simulations compare with observational constraints will be discussed in \sref{sec:discussion}.}
Moreover, the typical SFRs are somewhat less in \itng than in \illustris: at $z = 2$ and $\mstar \sim 10^{11} \, \msun$ (a representative stellar mass for bright SMGs),
the median SFR in \itng is $7 \, \msunperyr$, whereas the corresponding value for $\illustris$ is $63 \, \msunperyr$.
\cch{This difference is partially due to \itng having a lower normalisation for the $z \sim 2$ SFMS \citep{Donnari2019}
(and consequently a lower cosmic SFR density at `cosmic noon'; see fig. 4 of \citealt{Pillepich2018itng}) compared to \illustris
and partly because of the higher quenched fraction in this mass and redshift range in \itng.}
This indicates that one or more of the differences between the \illustris and \itng galaxy formation models results in systematically lower
star formation rates for massive $z \sim 2$ galaxies -- and consequently less SMGs -- in \itng compared with \illustris.

\begin{figure*}
\centering
\includegraphics[width = 0.48 \textwidth]{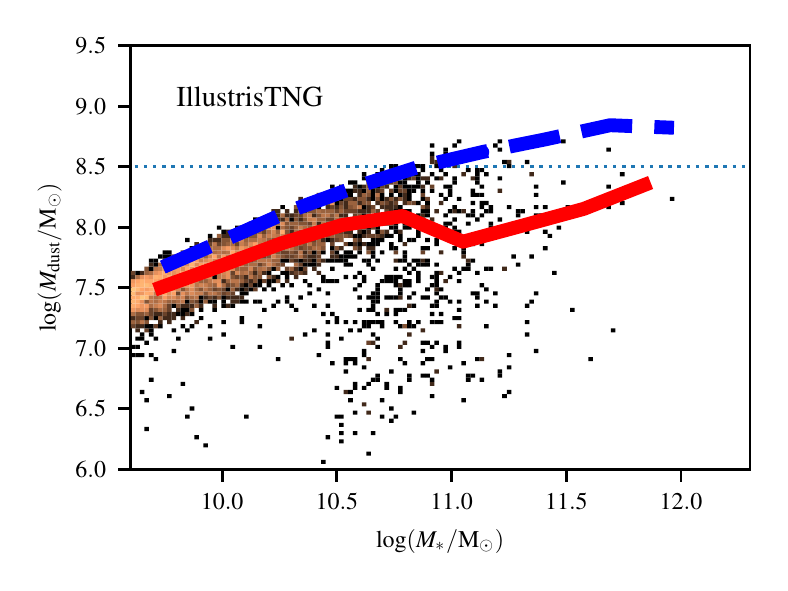}
\includegraphics[width = 0.48 \textwidth]{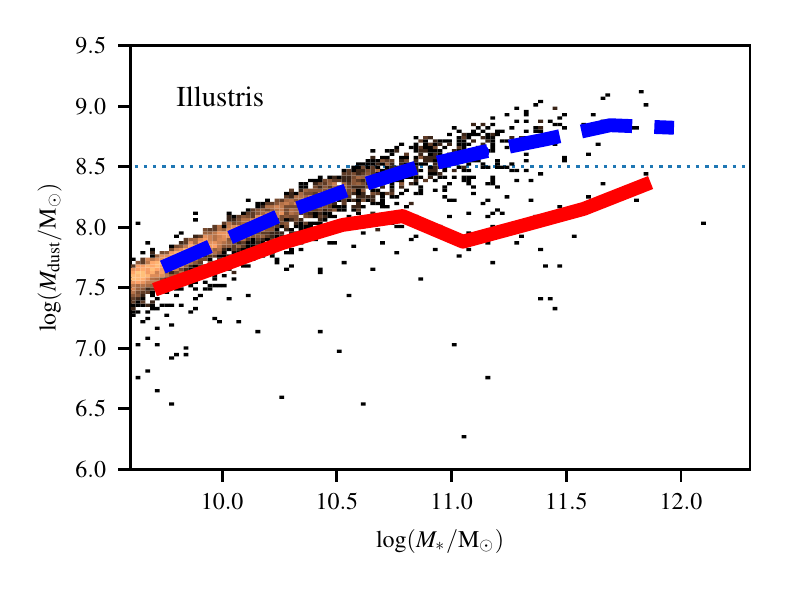}\\
\includegraphics[width = 0.48 \textwidth]{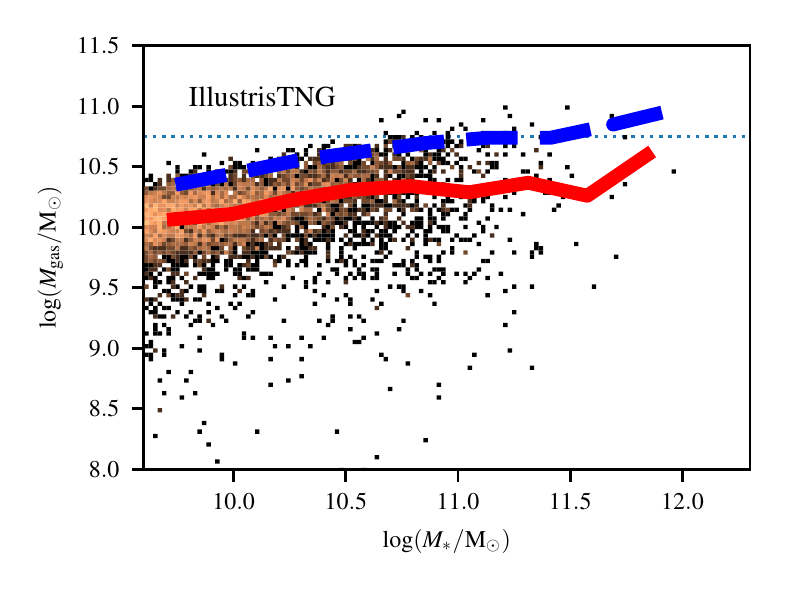}
\includegraphics[width = 0.48 \textwidth]{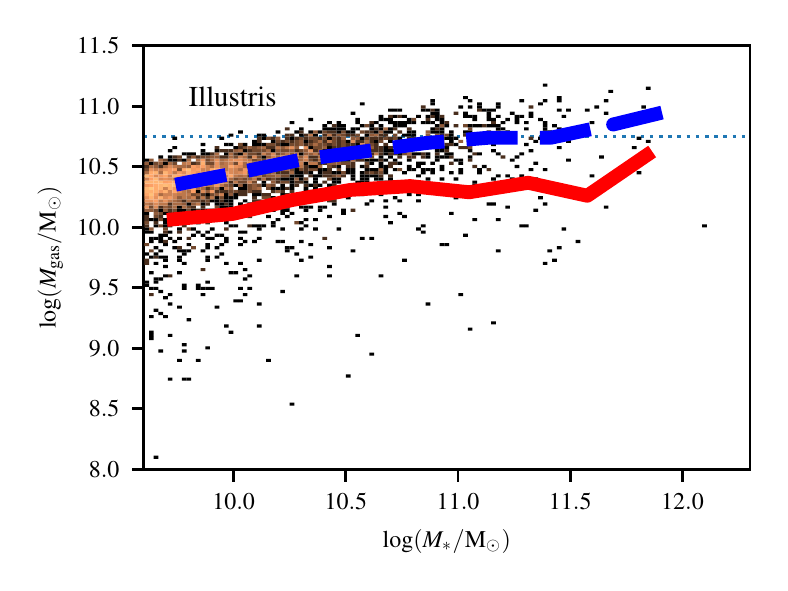}\\
\includegraphics[width = 0.48 \textwidth]{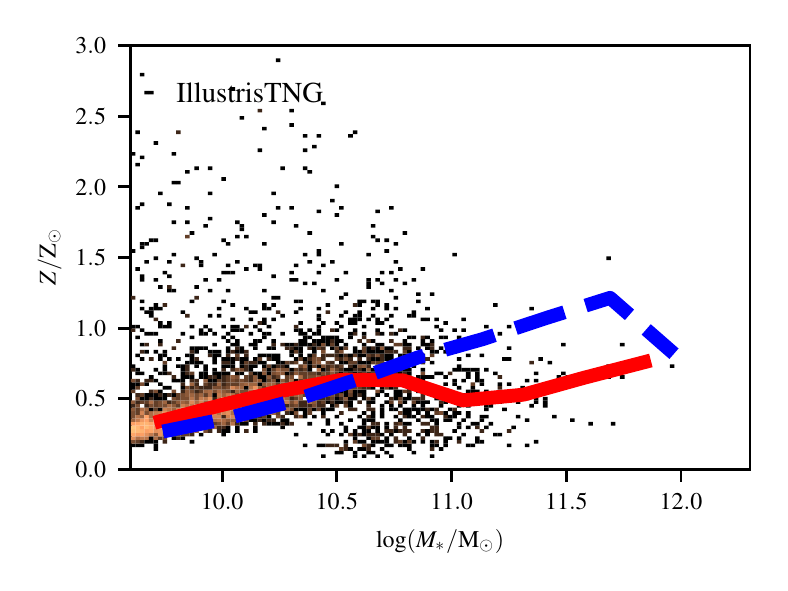}
\includegraphics[width = 0.48 \textwidth]{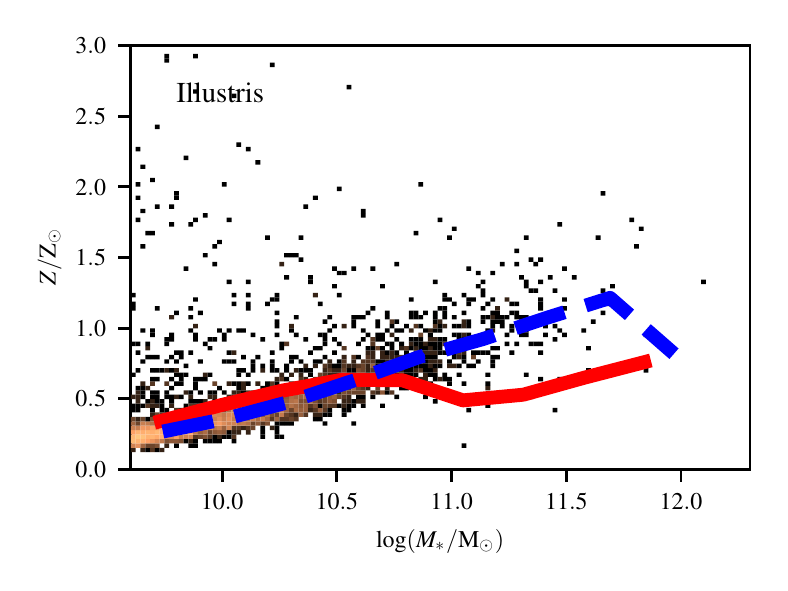}\\
\caption{Dust mass (\emph{first row}), gas mass (\emph{second row}), and metallicity (\emph{third row}) vs. stellar mass at $z=2$ for
\itng (\emph{left column}) and \illustris (\emph{right column}).
\rev{The red solid (blue dashed) lines indicate the running medians for \itng (\illustris).}
The horizontal lines are intended only to guide the eye. The \itng galaxies
have systematically lower dust masses than those in \illustris, especially at high stellar masses ($\mstar \ga 10^{10.5} \, \msun$).
This difference is primarily driven by the \itng galaxies having lower gas masses at fixed stellar mass. The normalisation of
the $z = 2$ mass--metallicity relation at $\mstar \ga 10^{10.5} \, \msun$ is also slightly lower in \itng than in \illustris, but in terms
of the effects on dust masses, the metallicity differences are subdominant to the differences in gas masses.}
\label{fig:mdust_props}
\end{figure*}

From \fref{sfr_mdust}, we saw that galaxies in \itng tend to have lower dust masses than those in \illustris. Recall that in our model, the dust masses of the simulated
galaxies are determined by their ISM cold gas-phase metal masses, so the differences in the $\mdust$ values must be driven by differences in gas masses or/and metallicities.
\fref{fig:mdust_props} shows dust mass (first row), gas mass (second row), and metallicity (third row) vs. stellar mass for \itng (left column) and \illustris (right column); all $z = 2$ galaxies
with $\mstar \ga 10^{9.5} \, \msun$ are plotted.
The first row of \fref{fig:mdust_props} shows that at $z = 2$ and $\mstar \ga 10^{9.5} \, \msun$, the \itng galaxies have systematically lower
dust masses (at fixed stellar mass) than the \illustris galaxies, and the difference is greater (a factor of a few) at $\mstar \ga 10^{10.5} \, \msun$.
At $z = 2$ and $\mstar \sim 10^{11} \, \msun$, the median $\mdust$ values in \itng and \illustris are $7.6 \times 10^7$ and $4.2 \times 10^8 \, \msun$, respectively, a factor of 3 difference.
The second row shows that this
discrepancy is primarily due to the \itng galaxies having lower gas masses than the \illustris galaxies at fixed stellar mass: 
at $z = 2$ and $\mstar \sim 10^{11} \, \msun$, the ISM gas masses are a factor of $\sim 2$ lower in \itng than in \illustris; the median values are $1.9 \times 10^{10}$ and $5.5 \times 10^{10} \, \msun$, respectively.
The normalisation of the $z = 2$ mass--metallicity
relation at $\mstar \ga 10^{10.5} \, \msun$ is also lower in \itng than in \illustris (third row), but the magnitude of this discrepancy is much less than for the gas mass. Again for $z = 2$ and $\mstar \sim 10^{11}
\msun$, the median $Z/{\rm Z}_{\odot}$ values for \itng and \illustris are 0.49 and 0.92, a difference of only $\sim 30$ per cent.
Consequently, the differences in gas mass are primarily responsible for the differences in dust mass between the two simulations.

\begin{figure}
\centering
\includegraphics[width = 0.48 \textwidth]{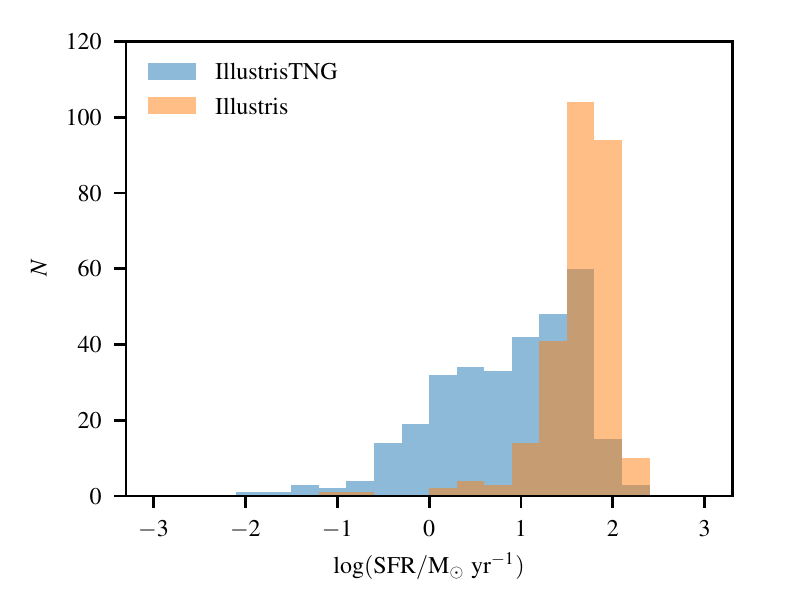}\\
\includegraphics[width = 0.48 \textwidth]{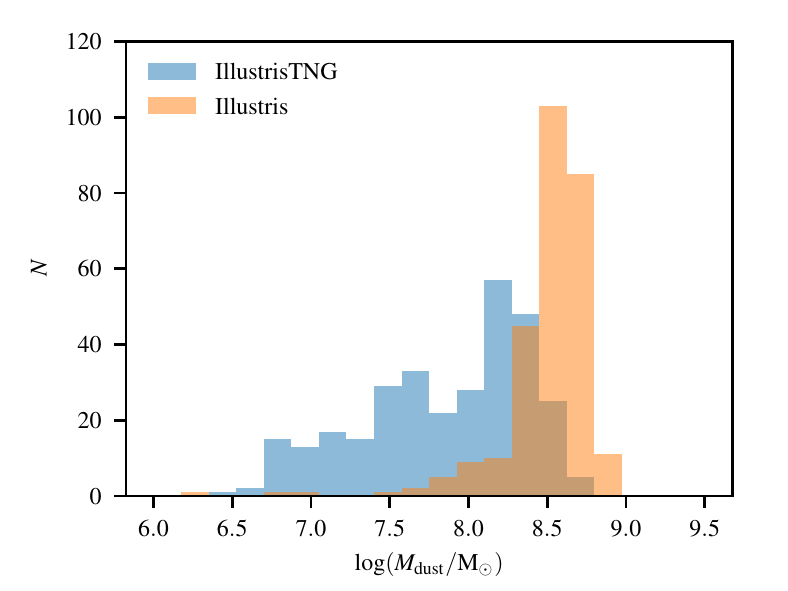}
\caption{SFR (\emph{top}) and dust mass (\emph{bottom}) for galaxies at $z = 2$ with $10.8 < \log (\mstar/\msun) < 11.2$
in \illustris (\emph{blue}; $N = 274$) and \itng (\emph{orange}; $N = 312$). For both SFR and dust mass, the \illustris distributions
are offset to higher values relative to the \itng distributions. SMGs tend to correspond to galaxies in the high-SFR and
high-$\mdust$ tails of the distributions. Consequently,
the fact that the highest SFR and $\mdust$ bins are significantly less populated in \itng than in \illustris causes the
former to host many fewer SMGs.
}
\label{fig:mstar11_comparison}
\end{figure}

To further highlight the differences between the simulations that drive the differences in their SMG populations, \fref{fig:mstar11_comparison} shows the distributions of SFR and $\mdust$
for galaxies at $z = 2$ with stellar masses of $\mstar \sim 10^{11} \msun$ $(10.8 < \log (\mstar/\msun) < 11.2)$.
Both the SFR (top panel) and dust mass (bottom panel) distributions of the $z = 2$ $\mstar \sim 10^{11} \msun$ galaxies in \itng are systematically offset to
lower values relative to those for \illustris galaxies. Even at this high stellar mass, the SMG selection tends to select only galaxies in the high-SFR or/and high-$\mdust$ tails of the distributions
(e.g. in our model, a $z = 2$ galaxy with SFR $= 100 \, \msunperyr$ and $\mdust = 10^{8.5} \, \msun$ has $S_{850} = 1.5$ mJy). Consequently, it is clear that the relative lack of SMGs in \itng compared
to \illustris occurs because there are significantly fewer galaxies in both the high-SFR and high-$\mdust$ tails of the distributions in \itng.

\begin{figure}
\centering
\includegraphics[width = 0.48 \textwidth]{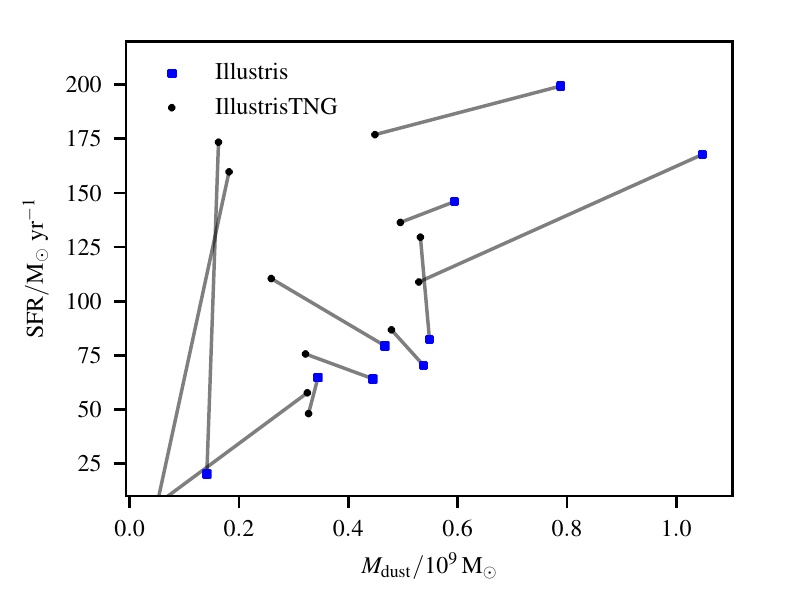}\\
\includegraphics[width = 0.48 \textwidth]{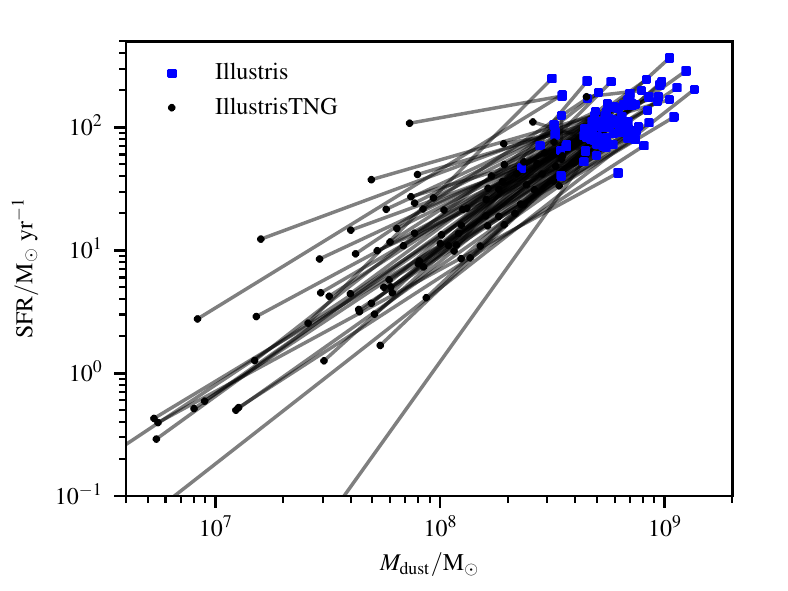}
\caption{\emph{Top: } The positions of the \rev{11} $z = 2$ SMGs with $S_{850} > 2$ mJy in \itng (blue points) in the SFR--$\mdust$ plane and the corresponding galaxies in \illustris (black points; lines connect points corresponding to the same halo
in different simulations). \emph{Bottom:} Similar, but for the 79 $z = 2$ SMGs with $S_{850} > 2$ mJy in \illustris. The top panel shows that the halos that host SMGs in \itng host massive galaxies with similar SFR and $\mdust$ (and
thus $S_{850}$) values in \illustris; in contrast, the bottom panel shows that most of the galaxies that are SMGs in \illustris have drastically reduced SFRs or/and dust masses in \itng, indicating that the differences in the feedback
models cause most of the \illustris SMGs to `drop out' of the \itng SMG population.}
\label{fig:sfrmdust_connected}
\end{figure}

\cch{Given that it would be prohibitively computationally expensive to run many variants of the simulations considered here,
systematically varying the sub-grid feedback models one-by-one, solely for the purpose of the present work,} it is somewhat challenging to identify which change(s) to the
feedback models cause \itng to host significantly fewer SMGs than \illustris. However, since the simulations use identical initial conditions, it is instructive to consider how the properties of matched individual galaxies differ between the two simulations.
For a given halo hosting an SMG with $S_{850} > 2$ mJy in \itng at a given redshift, we search \illustris for the halo at the same redshift whose center is closest to that of the \itng halo and confirm that the two halos' virial masses differ by less than 30 per cent;
we do the same for the halos that host SMGs in \illustris.
\fref{fig:sfrmdust_connected} presents this comparison for the SFR--$\mdust$ plane for SMGs in the $z = 2$ snapshot of \itng (top panel) and \illustris (bottom panel). The top panel shows the positions of the \itng SMGs in the
SFR--$\mdust$ plane as black points. Each of these galaxies is connected to the matched \illustris halo (blue points) by a line. The plot reveals that for all of the 6 $z = 2$ SMGs
in \itng, the corresponding galaxy in \illustris has similarly high SFR and $\mdust$ mass values. For individual galaxies, the median ratio of the stellar mass (not shown) in \illustris to that in \itng is 0.94. The median SFR ratio is
0.81, and the median dust mass ratio is 1.1. Consequently, halos that host SMGs in \itng tend to host galaxies with similarly high submm flux densities in \illustris.

The bottom panel of \fref{fig:sfrmdust_connected} shows a similar comparison but for the 79 $z = 2$ SMGs in \illustris. In contrast with the top panel, this plot
indicates that most of these galaxies have significantly lower SFR or/and $\mdust$ values in \itng. The median ratio of individual \illustris
SMGs' stellar masses (not shown) in \itng to the corresponding value for \illustris is 0.76. The median ratios for the SFR, dust mass, and gas mass (not shown) are 0.12, 0.17, and 0.24, respectively, indicating that halos that host SMGs
in \illustris typically host galaxies with order-of-magnitude-lower SFRs and dust masses in \itng.
Taken together, these plots demonstrate that some change(s) to the feedback model in \itng relative to the \illustris model have caused most of the SMGs in \illustris to `drop out' of the SMG population in \itng, primarily
by decreasing the gas mass and thus both SFR and dust mass. \cch{One of the primary drivers of the changes to the sub-grid models in \itng relative to those in \illustris was the need to quench galaxies more effectively.
Although this goal has been achieved in \itng in the sense that a $z = 0$ colour bimodality consistent with observations is obtained, it seems that the highest-mass galaxies are now being quenched slightly too early,
causing the SMG counts to be under-predicted by \itng.}

\section{Discussion} \label{sec:discussion}

\subsection{SMGs as a constraint on galaxy formation models}

We have shown that the SMG number counts predicted using \itng are significantly lower than observed, whereas those predicted using \illustris agree better with observations;
the reasons for this difference are that massive ($\mstar \sim 10^{11} \msun$) $z \sim 2$ galaxies in \itng tend to have significantly lower SFRs and dust masses
(primarily due to having lower gas masses) than in \illustris. This difference holds not only statistically but also for matched halos in the two simulations.
\cch{We conclude that one or more of the differences between the sub-grid feedback models employed
in \illustris and \itng~-- likely those for stellar feedback-driven outflows or/and AGN feedback --
are responsible for these differences.}

Overall, \itng is more successful than \illustris at matching various more-general observational constraints. In particular, multiple lines of evidence indicated that
the quenching of star formation in \illustris was too inefficient (see \citealt{Nelson2015illustris} for a detailed discussion), and the changes to the sub-grid feedback models were
partially motivated by the need to rectify this issue \citep{Weinberger2017,Weinberger2018,Nelson2018itng}. \cch{The models employed in \itng result in more-efficient quenching,
yielding a $z = 0$ bimodal colour distribution consistent with that observed. However, the resulting galaxy morphology--colour and morphology--size relations are inconsistent
with observations, suggesting that further refinement of the quenching model is necessary \citep{Rodriguez-Gomez2019}. Our results provide additional support for this,
suggesting that massive $z \sim 2$ galaxies are too efficiently quenched \cch{in \itng} (but see \citealt{Donnari2019} for a comparison of the quenched fractions at $z \sim 1.75$ in \illustris
and \itng with observational constraints). This may indicate that the \emph{timing} of quenching of massive galaxies is still not correct; we discuss
this issue in more detail in the next subsection.}

Our results reinforce the need to satisfy multiple observational constraints, such as the stellar mass function at different redshifts, before a model for the SMG population
can be considered successful. In their hybrid `semi-empirical' model, which obtained good agreement with observed SMG number counts,
\citet{HN13} partially avoided the issues encountered here by matching the stellar mass functions and
gas fractions of massive $z \sim 2$ galaxies \emph{by construction}; the `added value' was provided by including the effects of mergers based on hydrodynamical
simulations and performing radiative transfer to self-consistently compute submm fluxes. In contrast, models that have attempted to self-consistently reproduce
the SMG population in an `ab initio' manner have generally come up short because they do not reproduce the observed counts or other properties of SMGs, they
employ simplifications that are unlikely to be valid in detail, or they incorporate IMF variations that are not well-motivated by theory or observations.

For example, in the large-volume simulation of \citet{Dave:2010}, the `SMGs' (selected
assuming a monotonic mapping between SFR and submm flux, which should not hold in detail due to the influence of dust mass on the SED shape) have SFRs
a factor of $\sim 3$ less than observed. Also using a large-volume cosmological simulation, \citet{Shimizu:2012} predicted SMG number counts in reasonable
agreement with those observed, but, as discussed above, their simple model for computing the submm flux likely causes them to over-predict the submm
flux and thus counts compared with more accurate calculations.
Based on a single cosmological zoom simulation, \citet{Narayanan2015} claimed that
`typical' massive galaxies undergo a long-lived SMG phase and suggested that their model would predict counts consistent with those observed.
However, the central galaxy in their simulation is a bright SMG ($S_{850} > 5$ mJy) only once the dust mass exceeds $\sim5 \times 10^9 \msun$; this
value is greater than the dust masses inferred via SED modelling for any of the ALMA-resolved SMGs from \citet{daCunha2015} and \citet{Miettinen2017sed}.
Note that their simulation did not include AGN feedback, which is believed to be a crucial ingredient to match other observational constraints, and our
work suggests that the treatment of AGN feedback can significantly affect the predictions for the SMG population (see also \citealt{Lacey2015}).
Moreover, subsequent works employing an improved version of the stellar feedback model used in \citet{Narayanan2015} have found that massive galaxies
are insufficiently quenched \citep{Su2018cooling_flow_1,Su2020}. The lack of AGN feedback may explain why \citet{Narayanan2015} obtained
such high dust masses (up to $\sim 8 \times 10^{9} \msun$ for the central galaxy at the end of the simulation, $z = 2$).
The semi-analytic models that are able to match SMG number counts either employ top-heavy IMFs
\citep{Baugh:2005,Lacey2015} or `non-traditional' scalings for stellar feedback-driven outflows \citep{Safarzadeh2017}, and it remains unclear
whether these assumptions are consistent with other observational constraints and physical arguments.

Given the above, we conclude that producing SMGs in sufficient number with properties consistent with those observed remains a challenge for theorists,
and further work, both observational and theoretical, regarding this population is sorely needed. Because reproducing the SMG population in a cosmological
context is so difficult, they are useful for constraining sub-grid feedback models employed in cosmological simulations. With the exception of the {\sc galform}
SAM, SMGs are typically not considered when tuning existing sub-grid models or building new ones, perhaps because of the perceived complexity of
predicting SMG number counts (assuming that radiative transfer calculations are required). However, this can change. The simple method for predicting
submm flux that we have employed enables predicting SMG number counts from large-volume cosmological simulations with negligible additional computational expense.
Consequently, the observed SMG number counts and redshift distribution can be used as additional constraints on feedback models that are sensitive to regions
of the galaxy parameter space (namely, high mass, SFR, and redshift) that traditional comparisons tend not to probe.
\cch{The utility of SMGs for constraining galaxy formation physics will continue to increase as the community observes larger samples of submm sources with
ALMA to more accurately determine the number counts, especially at high flux densities and at high redshift.
One particularly interesting application will be constraining variations in the IMF (as has been the case for more than a decade): if the next generation of large-volume
simulations is able to reproduce many properties of the population of less extreme galaxies even better than is currently the case but still cannot
reproduce the SMG number counts and redshift distribution, this may indeed be (indirect) evidence for IMF variation in SMGs.}

\subsection{Quenched galaxies at $z \sim 2-3$ and their relationship to SMGs}

As discussed above, our results suggest that massive ($\mstar \sim 10^{11} \msun$) $z \sim 2-3$ galaxies are being quenched too early in \itng.
\rev{It should thus be instructive to compare the quenched fractions of massive $z \sim 2-3$ galaxies in the simulations with those observed.}
\citet{Donnari2019} show that the \itng quenched fraction at this mass at $z \la 1.75$ is consistent with the observations
of \citet{Muzzin2013}, of order 50 per cent. Moreover, \citet{Valentino2019} find that the number density of quenched galaxies with $\mstar \ga 4 \times 10^{10} \msun$
at $z \sim 3.5$ predicted by \itng is broadly consistent with observational constraints (which, however, are uncertain at the order-of-magnitude level).
\citet{Valentino2019} also show that the SMG number density at $z \ga 3$ in \itng predicted using our model is consistent with that observed.
\rev{However, neither of the aforementioned studies cover the peak of the observed SMG redshift distribution,
and $z \sim 2-3$ is the crucial period during which the massive galaxy population transitions from being dominated by dusty star-forming galaxies
to being dominated by quenched galaxies \citep[fig. 2 of][]{Martis2016}.
Using a significantly larger sample than previous studies, \citet{Sherman2020} find that \itng predicts quenched fractions for $\mstar \ge 10^{11}
\msun$ galaxies at $1.5 < z < 3$ that are $\sim 2-5$ times greater than observed. If indeed \itng overpredicts the quenched fraction as
suggested by \citet{Sherman2020}, we would naturally expect it to underpredict the observed SMG counts.}

\rev{Given the above results, it is plausible that massive galaxies are being quenched somewhat too early in \itng. However, the fact that the \itng
redshift distribution agrees well with that observed (see \fref{fig:z_dist}) seems inconsistent with this explanation.
Moreover, since it is known that \illustris underpredicts the quenched fraction and does not exhibit a $z \sim 0$ colour bimodality consistent
with observations, it is unsurprising that the \illustris SMG redshift distribution peaks at much lower redshift than observed.
Since \itng underpredicts the submm counts by more than a magnitude, we wish to avoid over-interpreting
the agreement in terms of the SMG redshift distribution, but these results clearly demonstrate that both the integral SMG number counts and
redshift distribution serve as useful constraints on galaxy formation models.}

\subsection{Limitations and future work}

\cch{We will now discuss some limitations and avenues for future work.
The simplicity of our model for predicting submm flux densities is an advantage in that it enables predicting SMG number counts from large-volume cosmological
simulations with negligible additional computational expense, but its simplicity also brings some limitations. First, the relation tends to over-predict
the submm flux densities of fainter (i.e. lower-SFR or/and lower-$\mdust$) SMGs because in such galaxies, the underlying assumption that all of the star formation
is obscured can break down. Thus, to apply Eq.~\ref{eq:s850} to such galaxies, it would be necessary to first estimate the fraction of the SFR that is obscured
by dust and then use that value rather than the total SFR (for one possible method to do so, see Popping et al., submitted).
Even better would be to perform dust radiative transfer directly on the simulated galaxies if we are convinced that the simulations resolve the clumpy
structure of the ISM sufficiently well to make this meaningful. A useful intermediate step would be to re-derive a scaling relation for submm flux density using
higher-resolution, more physical simulations than those originally used \citep{H11,HN13}. We have already performed this type of
verification on a limited set of simulations \citep{Liang2018,Cochrane2019}, but more work along these lines would be merited.}

\cch{Both the simple scaling relations we use and full dust radiative transfer calculations critically depend on the dust mass (specifically the dust mass distribution in the case of
the latter). In this work, we employ a cut in temperature-density space to define `ISM gas' and then multiply the total metal mass in this ISM
gas by a constant dust-to-metal ratio in order to approximate the dust mass. Using a `live' dust model in the simulations
\citep[e.g.][]{McKinnon2016,McKinnon2017} would potentially be a more accurate approach, especially since incorporation of dust destruction
processes would alleviate the need to identify dust-containing `ISM gas' with a simple cut like we have done here. Going further, treating the
dust as a separate fluid \cite[e.g.][]{Hopkins2016,Seligman2019,McKinnon2019} may lead to non-negligible differences in the results of radiative transfer calculations (and thus any
scaling relation we would derive from them) because e.g. if the dust and gas are not coupled, radiation-pressure-driven outflows may affect the
ISM gas and dust differently and may even lead to different grain properties as a function of distance from the galaxy.}

\cch{Moreover, we found in previous work that \illustris (and likely \itng given that it has the same resolution) has less starbursts (outliers above
the SFMS) than observed \citep{Sparre2015}. \citet{Sparre2017} demonstrated that resolution is one cause of the under-prediction
of starbursts in \illustris by performing cosmological zoom simulations of mergers selected from \illustris, keeping the sub-grid model fixed,
and finding order-of-magnitude higher SFR enhancements due to mergers in the zooms than in \illustris.
The fact that \illustris matches the observed SMG counts reasonably well suggests that if starbursts were better resolved, the \illustris sub-grid
model may actually lead to an over-prediction of the SMG counts. Conversely, the SMG counts predicted using the \itng model may be
in better agreement with observations if starbursts are resolved.}

\section{Conclusions} \label{sec:conclusions}

\cch{We have used a simple scaling relation derived in previous work} to predict SMG number counts for the \itng ~\cch{TNG100} and \illustris cosmological hydrodynamical
simulations. Our principal conclusions are the following:
\begin{enumerate}
\item The SMG number counts predicted based on \itng are \cch{lower than those observed for $S_{850} \ga 4$ mJy by an order of magnitude or more}; in contrast,
those based on \illustris are in excellent agreement with the observed counts for $S_{850} \la 4$ mJy and $S_{850} \ga 9$ mJy, and they are only a factor of $\sim 2$ lower than observed
for intermediate flux densities. Thus, \illustris fares better than \itng in terms of reproducing the observed SMG population \cch{(which mainly corresponds to galaxies
with $\mstar \ga 10^{10.5} \, \msun$ at $z \sim 2-4$)}
despite the latter better satisfying other
observational constraints, such as the stellar mass function at $z \la 1$, the $z = 0$ galaxy colour bimodality, the $z \la 2$ massive galaxy quenched fraction,
and the gas contents of low-redshift galaxies.

\item \rev{The redshift distribution of SMGs in \itng agrees well with that observed, with similar median redshifts (3.0 and 2.8, respectively), whereas that for
\illustris is biased toward significantly lower redshift than observed (median $z \sim 1.5$).}

\item The difference between the counts predicted based on the two simulations is due to $z \sim 2$ massive galaxies in \itng having both lower dust masses
(primarily because of lower gas masses rather than lower metallicities) and SFRs at fixed stellar mass compared to \illustris.

\item In both simulations, SMGs correspond to massive ($\mstar \sim 10^{11} \msun$) galaxies with SFR and dust mass values well above the median
value for galaxies at fixed mass and redshift. The high-$\mdust$ and, to a lesser extent, high-SFR tails of the distributions are more populated in \illustris
than in \itng, which causes the former to host considerably more SMGs.

\item By comparing the same halos in \illustris and \itng, we find that halos that host SMGs in \itng tend to host galaxies with similar SFR and $\mdust$
values in \illustris. However, the converse does not hold: most of the SMG-hosting halos in \illustris tend to host galaxies with similar stellar masses
but order-of-magnitude or more lower SFR and dust mass values in \itng.

\item Because the \illustris and \itng simulations employ the same initial conditions \cch{modulo minor differences in the cosmology},
this comparison indicates that the differences in the sub-grid models employed in the two simulations alter the predictions for the SMG population.
Overall, our results demonstrate that the SMG population provides important constraints
on galaxy formation models.

\cch{Our method entails negligible additional computational expense, so in
the future, the observed SMG number counts and redshift distribution can be employed as useful `orthogonal' constraints when developing and tuning
the sub-grid feedback models that are necessarily employed in large-volume cosmological simulations,
as has been done by some semi-analytic modelling groups for more than a decade.}

\end{enumerate}

\section*{Acknowledgements}

We thank Shy Genel, Melanie Habouzit, and Stuart McAlpine for useful discussion.
The Flatiron Institute is supported by the Simons Foundation.
MS acknowledges support from the European Research Council under ERC-CoG grant CRAGSMAN-646955.

\section*{Data Availability}

The data underlying this article will be shared on reasonable request to the corresponding author.

\bibliographystyle{mnras}
\bibliography{ref,smg,smg_obs,std_citations}

\bsp
\label{lastpage}

\end{document}